\documentclass[a4paper,11pt]{article}
\pdfoutput=1 

\usepackage{jheppub} 

\usepackage[T1]{fontenc} 

\title{A possible subthreshold pole in $S_{11}$ channel from $\pi N$ Roy-Steiner equation analyses}

\author[a]{Xiong-Hui Cao}
\author[a]{Qu-Zhi Li}

\author[b,1]{Han-Qing Zheng,\note{Corresponding author.}}
\affiliation[b]{%
 School of Physics and State Key Laboratory of Nuclear Physics and Technology, Peking University,\\
Beijing 100871, China
}%
\affiliation[b]{
 College of Physics, Sichuan University, Chengdu, Sichuan 610065, China
}%
\emailAdd{xionghuicao@pku.edu.cn}
\emailAdd{2001110075@stu.pku.edu.cn}
\emailAdd{zhenghq@scu.edu.cn }

\abstract{The hyperbolic version of Roy-Steiner equation describing low energy $\pi N$ scatterings, with larger analyticity domain in the complex $s$ plane is solved. 
The numerical results on phase shifts of low partial waves are in agreement with that of Hoferichter et al.~[\href{https://doi.org/10.1016/j.physrep.2016.02.002}{Phys.~Rept. 625 (2016) 1}].
A subthreshold pole in $S_{11}$ channel is found located at $\sqrt{s}=(918 \pm 3)-i (163 \pm 9)$~MeV. }

\begin{document} 
\maketitle
\flushbottom
\section{Introduction}

Modern interests in Steiner equation for $\pi N$ scattering~\cite{Baacke:1969fx, Baacke:1970mi, Steiner:1970mh, Steiner:1971ms, Hite:1973pm}\footnote{Steiner obtained a set of partial wave equations for $\pi N$ scattering based on fixed-$t$ DR$s$ in Refs.~\cite{Baacke:1969fx, Baacke:1970mi, Steiner:1970mh, Steiner:1971ms} three years before~\cite{Roy:1971tc}, and also obtained in Ref.~\cite{Hite:1973pm} a set of partial wave equations based on hyperbolic DR$s$ (i.e. on hyperbolas in the Mandelstam plane). But following the common usage in the literature, we sometimes still call Steiner equation by Roy-Steiner (RS) equation.} and Roy equation for $\pi\pi$ scattering~\cite{Roy:1971tc} are initiated by the remarkable analyses on $\pi\pi$ scatterings~\cite{Ananthanarayan:2000ht, Caprini:2005zr, Kaminski:2006qe, Moussallam:2011zg, Caprini:2011ky}, and $\pi K$ scatterings~\cite{Buettiker:2003pp, Descotes-Genon:2006sdr, Pelaez:2018qny}, aiming at producing precisely low energy phase shifts, and the pole locations of $\sigma/f_0(500)$ and $\kappa/K^*_0(700)$.  Especially, for the latter purpose,  in  the case of unequal mass $\pi K$ scatterings, an extension  of analyticity domain in the complex $s$ plane is required by recalling the hyperbolic dispersion relations (DR$s$) derived in \cite{Hite:1973pm}, see also \cite{Descotes-Genon:2006sdr}.
The $\pi N$ scattering processes have also been extensively studied~\cite{Ditsche:2012fv, Hoferichter:2015dsa, Hoferichter:2015tha, Hoferichter:2015hva}, and low energy phase shifts as well as other low energy parameters are determined with impressive high accuracy. Nevertheless for $\pi N$ scatterings it lacks of  an expedition in the complex $s$ plane, unlike what has been done for the $\pi K$ scattering case~\cite{Descotes-Genon:2006sdr, Pelaez:2020uiw, Pelaez:2020gnd}. 
However, there is a strong motivation for the exploration to the complex $s$ plane~\cite{Wang:2017agd, Li:2021tnt, Chen:2022zgm}, where the existence of a subthreshold pole, is suggested in the $S_{11}$ amplitude.

Roy and RS equations consist of a group of equations for partial waves (PW$s$) that respect analyticity, unitarity, and crossing symmetry of S matrix.
In this work, the existence of the subthreshold pole can be proven by   RS equations which fulfil all symmetries and  whose inputs are only the available experimental data.
We will rely on  for $\pi N\ S_{11}$ amplitude $f^{1/2}_{0+}$ without further theoretical approximation~\cite{Ditsche:2012fv}:
\begin{align}\label{eq:f}
    f_{0+}^{1 / 2}(W)=& N_{0+}^{1 / 2}(W)\nonumber\\
    &+\frac{1}{\pi} \int_{W_{+}}^{\infty} \mathrm{d}W^{\prime} \sum_{\ell^{\prime}=0}^{\infty} \frac{1}{3}\left\{K_{0 \ell^{\prime}}^{1 / 2}\left(W, W^{\prime}\right) \operatorname{Im} f_{\ell^{\prime}+}^{1 / 2}\left(W^{\prime}\right)+2 K_{0 \ell^{\prime}}^{3 / 2}\left(W, W^{\prime}\right) \operatorname{Im} f_{\ell^{\prime}+}^{3 / 2}\left(W^{\prime}\right)\right. \nonumber\\
    &\left.+K_{0 \ell^{\prime}}^{1 / 2}\left(W,-W^{\prime}\right) \operatorname{Im} f_{\left(\ell^{\prime}+1\right)-}^{1 / 2}\left(W, W^{\prime}\right)+2 K_{0 \ell^{\prime}}^{3 / 2}\left(W,-W^{\prime}\right) \operatorname{Im} f_{\left(\ell^{\prime}+1\right)-}^{3 / 2}\left(W^{\prime}\right)\right\} \nonumber\\
    &+\frac{1}{\pi} \int_{t_{\pi}}^{\infty} \mathrm{d} t^{\prime} \sum_{J=0}^{\infty} \frac{\left(3-(-1)^{J}\right)}{2}\left\{G_{0 J}\left(W, t^{\prime}\right) \operatorname{Im} f_{+}^{J}\left(t^{\prime}\right)+H_{0 J}\left(W, t^{\prime}\right) \operatorname{Im} f_{-}^{J}\left(t^{\prime}\right)\right\}\ ,
\end{align}
where $W_+=m_\pi+m_N,t_\pi=4m_\pi^2$; $N^{1/2}_{0+}$ denotes nucleon pole term;  $K_{\ell \ell^\prime}$ are the kernels which act on the $\pi N \to \pi N$ PW$s$ $f^{I}_{\ell\pm}$, and $G_{\ell J},H_{\ell J}$ are the kernels which act on the $\pi\pi \to N\bar{N}$ PW$s$ $f^{J}_{\pm}$.
Detailed expressions for those kernels can be found in Ref.~\cite{Hoferichter:2015hva}.
RS equations hold for all PW$s$, but those for the $S$ and $P$ waves embody most of the characteristics of elastic $\pi N$ scatterings.
Previous works on RS equations of $\pi N$ scatterings concerned the properties at low energies on the real $s$ axis, such as $\pi N\ \sigma$ term~\cite{Hoferichter:2015dsa} and chiral low energy constants~\cite{Hoferichter:2015tha} etc.
The purpose of this work is, however, to study the analytic structure of PW amplitude in the complex $s$ plane.
Hence previous work has to be extended to accommodate for the new situation.

The present work, ``standing on the shoulder of Ref.~\cite{Hoferichter:2015hva}'', follows the spirit of Ref.~\cite{Descotes-Genon:2006sdr} in dealing with exotic resonance $\kappa/K^*_0(700)$.
We show how to extend the validity domain of $\pi N$ RS equations to the complex $s$ plane and use this extension to prove the existence of a second sheet subthreshold pole in $S_{11}$ channel and determine the position of this pole within small uncertainties.

\section{Two dispersion representations and their validity domains}

\begin{figure}[h]
\centering
    \includegraphics[width=8cm]{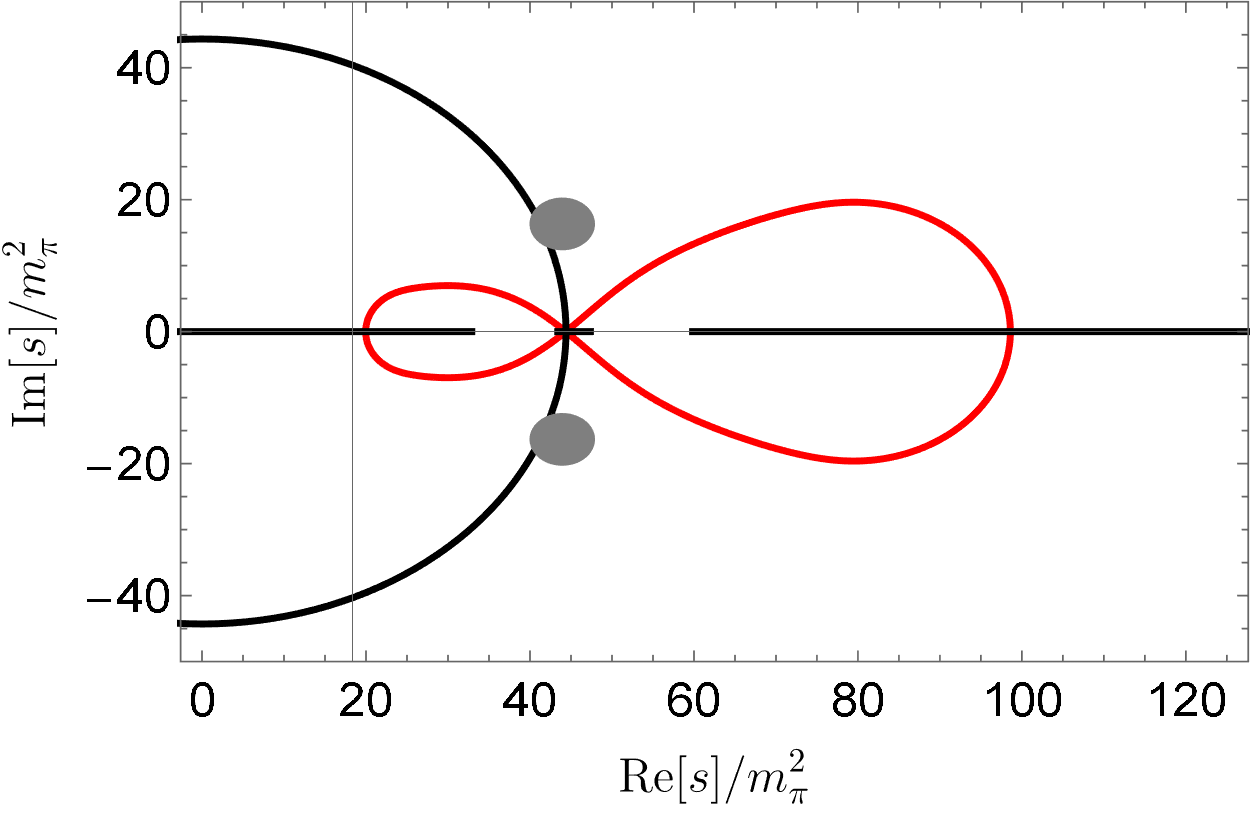}
    \caption{Validity domain of the fixed-$t$ representation~\cite{Baacke:1970mi}. The three cuts along the real axis as well as the circular cut of PW amplitude are also shown.
    The gray region indicates the possible position of the subthreshold pole.}\label{fig:fix_t}
\end{figure}

To begin with, we first introduce the traditional fixed-$t$ dispersion representation~\cite{Baacke:1970mi, Steiner:1970mh, Steiner:1971ms}. 
From axiomatic field theory~\cite{Martin:1969ina, Sommer:1970mr}, fixed-$t$ DR can be shown to be valid in a finite region of $t$.
The validity range relies on the convergence of the PW expansion of imaginary part of $\pi N$ amplitudes.
Lehmann~\cite{Lehmann:1958ita}  proved that the region is constrained by an ellipse with foci at $t=0$ and $t=-\lambda_{s^\prime}/s^{\prime}$, where $\lambda_{s^\prime}\equiv\lambda(s^{\prime},m_N^2,m_\pi^2)=(s^{\prime}-s_-)(s^{\prime}-s_+)$, $s_{\pm}=(m_{\pi} \pm m_N)^2$. The right extremity of the ellipse is at $t=T(s^\prime)$, where $s^\prime \geq s_+$ is the variable of dispersive integration.
We assume that the scattering amplitudes satisfy Mandelstam double spectral representation~\cite{Mandelstam:1958xc, Mandelstam:1959bc}, so the extremity of the ellipse is limited by the singularities from double spectral functions $\rho_{st},\rho_{su}$.
It may be written as $t=T_{st}(s)$ with $T_{s t}(s)=\min \left\{T_{\mathrm{I}}(s), T_{\mathrm{II}}(s)\right\}$, where~\cite{Mandelstam:1958xc}
\begin{align}
    \begin{aligned}
    &T_{\mathrm{I}}(s)=\frac{4 m_{\pi}^{2}\left(s-m_{N}^{2}-2 m_{\pi}^{2}\right)^{2}}{\lambda\left(s, m_{N}^{2}, 4 m_{\pi}^{2}\right)}>4 m_{\pi}^{2}\ , \quad \forall s>\left(m_{N}+2 m_{\pi}\right)^{2}\ , \\
    &T_{\mathrm{II}}(s)=\frac{16 m_{\pi}^{2}\left(s-m_N^2+m_\pi^2\right)^{2}}{\lambda\left(s, m_{N}^{2}, m_{\pi}^{2}\right)}>16 m_{\pi}^{2}\ , \quad \forall s>s_{+}\ .
    \end{aligned}
\end{align}
The expression of the boundary corresponding to $\rho_{su}$ can be put in the form $t=T_{su}(s)$ with $T_{s u}(s)=\max \left\{T_{\mathrm{III}}(t), T_{\mathrm{IV}}(t)\right\}$, where the boundary functions $T_{\mathrm{III},\mathrm{IV}}$ obey respectively~\cite{Frazer:1960zza}, 
\begin{align}
    \begin{aligned}
    &\lambda\left(u, m_{N}^{2}, m_{\pi}^{2}\right) \lambda\left(s, m_{N}^{2}, 4 m_{\pi}^{2}\right)-16 m_{\pi}^{2}\left[m_{N}^{2} s u-(m_\pi^2-m_N^2)^{2}\left(m_{N}^{2}-T_{\mathrm{III}}(s)\right)\right]=0\ , \\
    &\lambda\left(s, m_{N}^{2}, m_{\pi}^{2}\right) \lambda\left(u, m_{N}^{2}, 4 m_{\pi}^{2}\right)-16 m_{\pi}^{2}\left[m_{N}^{2} s u-(m_\pi^2-m_N^2)^{2}\left(m_{N}^{2}-T_{\mathrm{IV}}(s)\right)\right]=0\ ,
    \end{aligned}
\end{align}
where $u=2(m_\pi^2+m_N^2)-s-T_{\mathrm{III},\mathrm{IV}}(s)$.
Generally, $T_{st}$ gives  stronger restrictions to the validity domain.
The boundary of the domain can be expressed in polar coordinates on the complex $t$ plane~\cite{Descotes-Genon:2006sdr}:
\begin{align}
    T(\theta)=\min _{s^{\prime} \geq s_+} T\left(s^{\prime}, \theta\right)\ ,\quad T\left(s^{\prime}, \theta\right)=\frac{T_{s t}\left(s^{\prime}\right)\left(\lambda_{s^{\prime}}+s^{\prime} T_{s t}\left(s^{\prime}\right)\right)}{\lambda_{s^{\prime}} \cos ^{2} \frac{\theta}{2}+s^{\prime} T_{s t}\left(s^{\prime}\right)}\ .
\end{align}
The PW projection involves the values of the amplitude on the segment $t \in [-\lambda_s/s,0]$.
The boundary of the validity domain of the fixed-$t$ representation in the $s$ plane is therefore obtained by solving $\lambda_{s}+s T(\theta) \exp (i \theta)=0$.
The result is displayed in Fig.~\ref{fig:fix_t}.
Unfortunately, the possible position of the subthreshold pole is outside the validity region.
Hence one has to go beyond the fixed-$t$ DR and to proceed with the hyperbolic DR~\cite{Hite:1973pm}.

The hyperbolic DR is restricted in a hyperbola $(s-a)(u-a)=b$, where $a,b$ are hyperbolic parameters.
The $\pi N$ scattering is described by four different amplitudes, $A^{+}(s,t),B^{-}(s,t)$ and $A^{-}(s,t),B^{+}(s,t)$, which are even and odd respectively under $s,u$ exchange~\cite{Hohler:1983}.
We focus on, e. g., $A^+(s,t)$ and write a hyperbolic DR without subtraction~\cite{Hite:1973pm}:
\begin{align}
    A^{+}(s, t)=\frac{1}{\pi} \int_{s_{+}}^{\infty} \mathrm{d} s^{\prime}\left(\frac{1}{s^{\prime}-s}+\frac{1}{s^{\prime}-u}-\frac{1}{s^{\prime}-a}\right) \operatorname{Im}_s A^{+}\left(s^{\prime}, t^{\prime}\right)+\frac{1}{\pi} \int_{t_{\pi}}^{\infty} \mathrm{d} t^{\prime} \frac{\operatorname{Im}_t A^{+}\left(s^{\prime}, t^{\prime}\right)}{t^{\prime}-t}\ ,
\end{align}
and $\operatorname{Im}_{s,t}$ are proportional to the discontinuity along the $s,t$ cuts.
According to the fact that the internal variables satisfy $(s^\prime-a)(u^\prime-a)=b$, the domain of $s^\prime$ and $t^\prime$ can eventually pass to the parameter $b$.
We denote by $B_{s/t}(s^\prime/t^\prime,\theta)$ the description for $b$ in polar coordinates, which satisfy respectively~\cite{Ditsche:2012fv},
\begin{align}
    \begin{aligned}\label{eq:B(theta)}
    &\frac{\left(1+\frac{2 s^{\prime}}{\lambda_{s^{\prime}}}\left(\Sigma-s^{\prime}-a-\frac{B_{s}\left(s^{\prime}, \theta\right) \cos \theta}{s^{\prime}-a}\right)\right)^{2}}{A_{s^{\prime}}^{2}}+\frac{\left(\frac{2 s^{\prime}}{\lambda_{s^{\prime}}}\left(\frac{B_{s}\left(s^{\prime}, \theta\right) \sin \theta}{s^{\prime}-a}\right)\right)^{2}}{A_{s^{\prime}}^{2}-1} =1\ ,\\
    &\frac{\left(\frac{\left(t^{\prime}-\Sigma+2 a\right)^{2}-4 B_{t}\left(t^{\prime}, \theta\right) \cos \theta}{(t^\prime-t_\pi)(t^\prime-t_N)}-\frac{1}{2}\right)^{2}}{A_{t^{\prime}}^{2}-\frac{1}{2}}+\frac{\left(\frac{4 B_{t}\left(t^{\prime}, \theta\right) \sin \theta}{(t^\prime-t_\pi)(t^\prime-t_N)}\right)^{2}}{A_{t^\prime}\sqrt{A_{t^{\prime}}^{2}-1}} =1\ ,
    \end{aligned}
\end{align}
where $\Sigma=2(m_\pi^2+m_N^2), t_N=4 m_N^2, A_{s^\prime}=1+\frac{2 s^{\prime} T_{s t}\left(s^{\prime}\right)}{\lambda_{s^{\prime}}}\ (s^{\prime}>s_{+})$ and $A_{t^\prime}^2=\frac{16 m_{N}^2 N_{s t}\left(t^{\prime}\right)}{(t^\prime-t_\pi)(t^\prime-t_N)}\ (t^{\prime}>t_{N}), A_{t^\prime}^2=1-\frac{16 m_{N}^2 N_{s t}\left(t^{\prime}\right)}{(t^\prime-t_\pi)(t-t_N)}\ (t_N>t^{\prime}>t_{\pi})$.
The details of boundary function $N_{st}(t^\prime)$ is referred to Ref.~\cite{Ditsche:2012fv}.
The domain of validity for parameter $b$ is defined as $B_{s}(\theta)=\min _{m_{+}^{2} \leq s^{\prime}} B_{s}\left(s^{\prime}, \theta\right)$ and $B_{t}(\theta)=\min _{t_{\pi} \leq t^{\prime}} B_{t}\left(t^{\prime}, \theta\right)$.
The discontinuity functions $\operatorname{Im}_s A^+(s^\prime,t^\prime)$ and $\operatorname{Im}_t A^+(s^\prime,t^\prime)$ must be defined inside the $s^\prime$ and $t^\prime$ integration regions, once these functions are expanded on $\pi N \to \pi N$ and $\pi\pi \to N\bar{N}$ PW$s$, respectively.
Therefore the segment of PW integration via parameter $b$ [i.e., the end points at $(s-a)((m_{N}^{2}-m_{\pi}^{2})^{2} / s-a)\ \text{and}\ (s-a)(\Sigma-s-a)$], must be inside the validity domain of $b$.
As indicated above, the boundary in the $s$ plane for hyperbolic representation is obtained by
\begin{align}
    \begin{aligned}\label{eq:RS_b}
    &(s-a)(\Sigma-s-a)-B_{s/t}(\theta) \exp (i \theta) =0\ , \\
    &(s-a)\left(\left(m_{N}^{2}-m_{\pi}^{2}\right)^{2} / s-a\right)-B_{s/t}(\theta) \exp (i \theta) =0\ .
    \end{aligned}
\end{align}
Corresponding results of $\rho_{su}$ are easy to obtain once replacing $T_{st}$ with $T_{su}$ etc.

\begin{figure}[h]
    \centering
	\includegraphics[width=8cm]{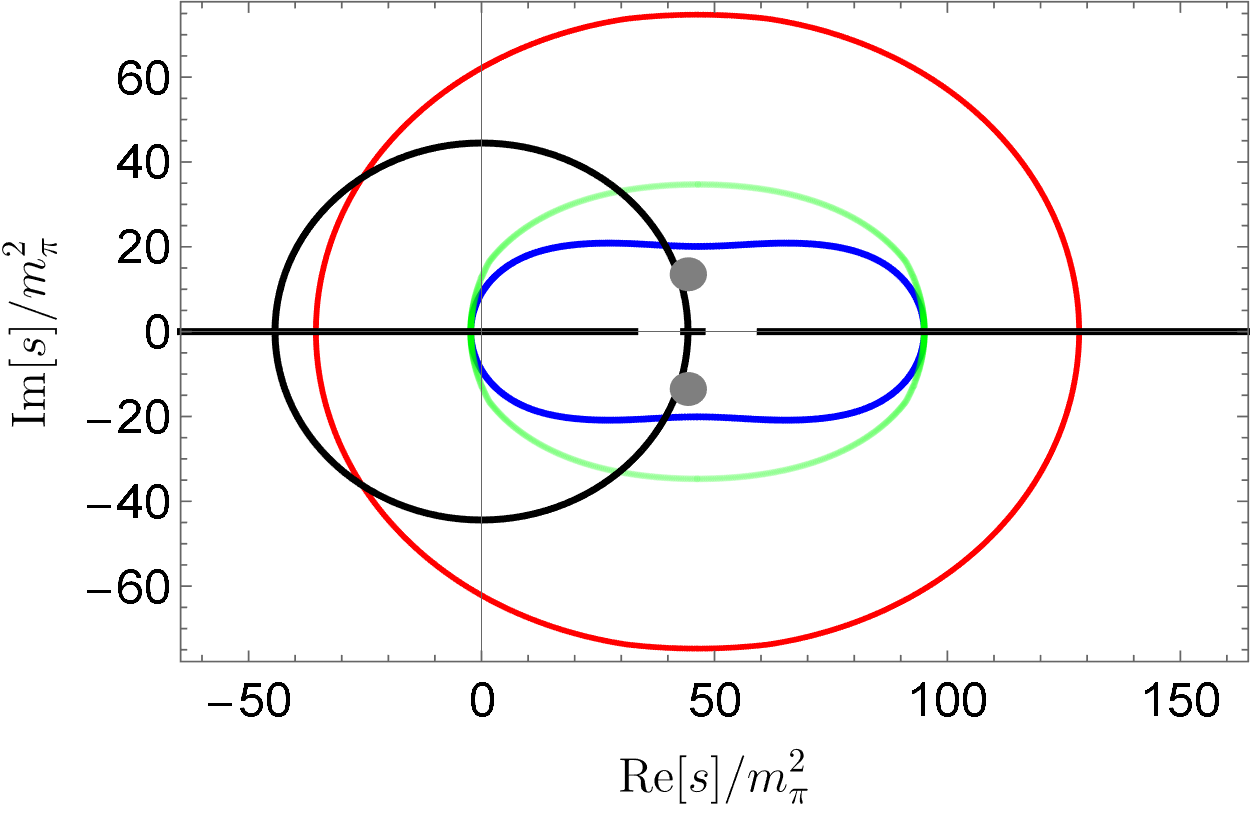}
	\caption{Validity domain of the fixed-$b$ RS representation ($a=0$).
	The blue and green lines correspond to the boundaries in the $s^\prime$ and $t^\prime$ integrals associated with $\rho_{st}$, respectively.
	The red line corresponds to the boundaries in the $s^\prime$ integral associated with $\rho_{su}$.
	}
    \label{fig:fixb_PiN}
\end{figure}

There is one comment worth noting.
In Eq.~(\ref{eq:B(theta)}), the polar radius $B_{s/t}(s^\prime/t^\prime,\theta)$ must be positive.
Eq.~(\ref{eq:B(theta)}) are quadratic equations of $B_{s/t}(s^\prime/t^\prime,\theta)$, thus the discriminant of the quadratic equations should be nonnegative.
Therefore the strongest restriction is $-2.59 m_{\pi}^2\ (\simeq -0.051~\mathrm{GeV}^2) < a < 4 m_{\pi}^2$.
If and only if in this region of $a$, the hyperbolic RS representation can be analytically extended to the complex $s$ plane. 
In Ref.~\cite{Hoferichter:2015hva}, it is chosen that $a=-29.3 m_\pi^2 (<-2.59 m_\pi^2)$, thus the solution of PW$s$ cannot be extended to the complex $s$ plane, although the solution has the largest range of validity on the real axis.
The validity domain derived from $s^\prime$ and $t^\prime$ integrals are displayed in Fig.~\ref{fig:fixb_PiN}.
This domain is larger along the imaginary direction than that of the Fig.~\ref{fig:fix_t}.
In the following, we will always set $a=0$, thus the largest validity value of $\sqrt{s}$ is $W_\mathrm{m}=1.36~\mathrm{GeV}$, slightly smaller than $1.38~\mathrm{GeV}$ in Ref.~\cite{Hoferichter:2015hva}.
Indeed, our domain sufficiently covers the possible position of the subthreshold pole.

\section{Solving the $s$-channel problem}

For our analyses, it is essential that the dispersion integral is dominated by the contributions from the low energy region and the PW expansion is useful only at low energies.
Therefore, the RS representation in Eq.~(\ref{eq:f}) is truncated to  include only $S$ and $P$ waves and we only solve such waves up to the matching point $W_\mathrm{m}=1.36~\mathrm{GeV}$.
The phase shift inputs at this point are taken from solutions of \cite{Hoferichter:2015hva}.
The $D$ and $F$ waves and the intermediate energy contributions of $\ell\leq 4$ PW$s$ are taken from GWU/SAID experimental data~\cite{Workman:2012hx} up to $W=2.5~\mathrm{GeV}$.
All high PW$s$ ($\ell>4$) and high energy tails can be estimated by Regge theory and are negligible~\cite{Hoferichter:2015hva}.
The price of using RS equations is
that it also requires input from the crossed channel ($t$-channel), whose PW$s$ $f^J_{\pm}$ have the similar expressions of Eq.~(\ref{eq:f})~\cite{Hoferichter:2015hva}.
As for the $t$-channel problem, the largest value of validity is $\sqrt{t}=1.94~\mathrm{GeV}$, so setting the matching point of $t$-channel, i.e., $t_\mathrm{m}=4m_N^2$ is reasonable.  
However, the $t$-channel imaginary parts $\operatorname{Im}f^J_{\pm}$ are available only above the two-nucleon threshold.
The amplitudes in the pseudophysical region $t_\pi \leq t \leq t_N$ need to be constructed from unitarity, or more precisely, Muskhelishvili-Omn\`es (MO) solutions~\cite{Omnes:1958hv, Muskhelishvili:1953}.
Due to the strong coupling of $K\bar{K}$ intermediate states to $f_0(980)$ in $S$ wave of $\pi\pi \to N\bar{N}$ scatterings, we follow the method~\cite{Moussallam:1999aq, Hoferichter:2012wf, Yao:2018tqn} and adopt a couple channel MO framework for the $\pi\pi, K\bar{K}$ system.
As for $P$ and $D$ waves, single channel approximation is sufficient in RS representation~\cite{Hoferichter:2015hva}. 
For more details, we refer to Refs.~\cite{Hoferichter:2012wf,Hoferichter:2015hva} about $t$-channel MO solutions. By using which, the RS problem which mixes $s$- and $t$-channel PW$s$ can be recasted as a Roy-like problem.

The mathematical properties of Roy (-like) equations, as a group of infinite coupled integral equations, have been thoroughly investigated in Refs.~\cite{Epele:1977um, Epele:1977un, Gasser:1999hz, Wanders:2000mn}.
In $\pi N$ RS analyses, we focus on a solution of the $S$ and $P$ waves of $s$-channel, within matching point $W_\mathrm{m}$.
Therefore the multiplicity of the coupled integral equations is $m=-2$~\cite{Hoferichter:2015hva}.
In addition, there are six ``no cusp'' constraints for phase shifts at the matching points.
Near the threshold, the $S$ wave scattering lengths were fixed precisely by the pionic atom spectrum results~\cite{Baru:2010xn},
\begin{align}
    a_{0+}^{1 / 2}=(169.8 \pm 2.0) \times 10^{-3} m_{\pi}^{-1}\ ,\quad a_{0+}^{3 / 2}=(-86.3 \pm 1.8) \times 10^{-3} m_{\pi}^{-1}\ ,
\end{align}
which serve as two additional constraints.
The mathematical theory of integral equations show that this corresponds to a system of $|m|+6+2=10$ constraints~\cite{Muskhelishvili:1953, Wanders:2000mn}.
As such, a unique solution exists if and only if the system also has ten free parameters.
For these reasons, it is convenient to take  subthreshold subtractions at $(\nu\equiv\frac{s-u}{4m_N}=0,t=0)$ and introduce ten subthreshold constants to match the number of degrees of freedom of the $\pi N$ RS system (see \cite{Hoferichter:2015hva} for more details).

\begin{figure}[h]
    \center
    \includegraphics[width=12cm]{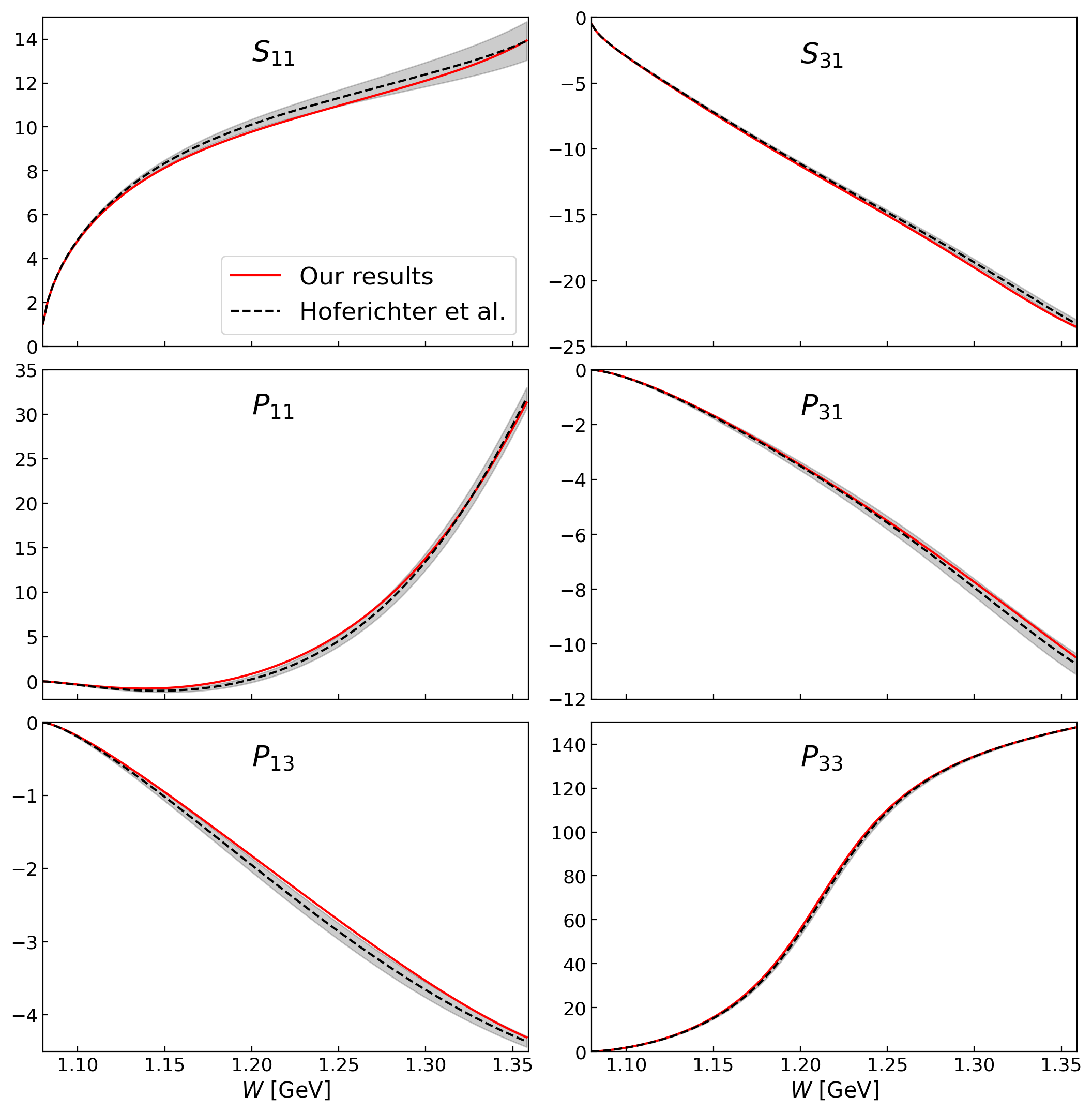}
    \caption{Phase shifts of the $s$-channel PW$s$ from our solutions (solid line) and \cite{Hoferichter:2015hva} (dashed line with error bands) in the low-energy region.
    }\label{fig:phase_6}
\end{figure}

Following \cite{Ananthanarayan:2000ht, Buettiker:2003pp, Hoferichter:2015hva}, we pursue the strategy in getting the $s$-channel solution: the phase shifts are each parameterized in a convenient way with a few parameters, which are matched to input PW$s$ above $W_\mathrm{m}$ in a smooth way.
To obtain an available solution of RS equations, a $\chi^2$-like function is introduced as,
\begin{align}
    \chi^{2}=\sum_{\ell, I, \pm} \sum_{j=1}^{N}\left(\frac{\operatorname{Re} f_{\ell \pm}^{I}\left(W_{j}\right)-F\left[f_{\ell \pm}^{I}\right]\left(W_{j}\right)}{\operatorname{Re} f_{\ell \pm}^{I}\left(W_{j}\right)}\right)^{2}\ ,
\end{align}
where $\{W_j\}$ denotes a set of points between threshold and matching point $W_\mathrm{m}$, and $F[f^{I}_{\ell\pm}]$ denotes the right-hand side of the RS equations.
The details of the technique have been explained in Ref.~\cite{Hoferichter:2015hva}.
By virtue of MO solutions where the phase-shifts and the moduli of $\pi\pi \to K \bar{K}$ $S$ wave amplitude from Refs.~\cite{Garcia-Martin:2011iqs, Niecknig:2012sj, Pelaez:2018qny} as $t$-channel inputs, we can obtain a coupled equations in terms of $s$-channel $S$ and $P$ waves.
The final numerical solution is equivalent to finding a minimized $\chi^2$ in the parameter space of the subtraction constants and the parameters describing the low-energy phase shifts.
Our solutions of phase shifts depicted in Fig.~\ref{fig:phase_6} are close to results of Ref.~\cite{Hoferichter:2015hva} within uncertainties and verifies the robustness of the RS method.
Here we haven't given the error analyses of the phase shifts, since our main focus is to verify whether there exists a subthreshold pole. 
However we do give an error analysis on the pole location in the following.

\section{The subthreshold singularities in $S_{11}$ channel}

\begin{figure}[h]
    \includegraphics[width=15cm]{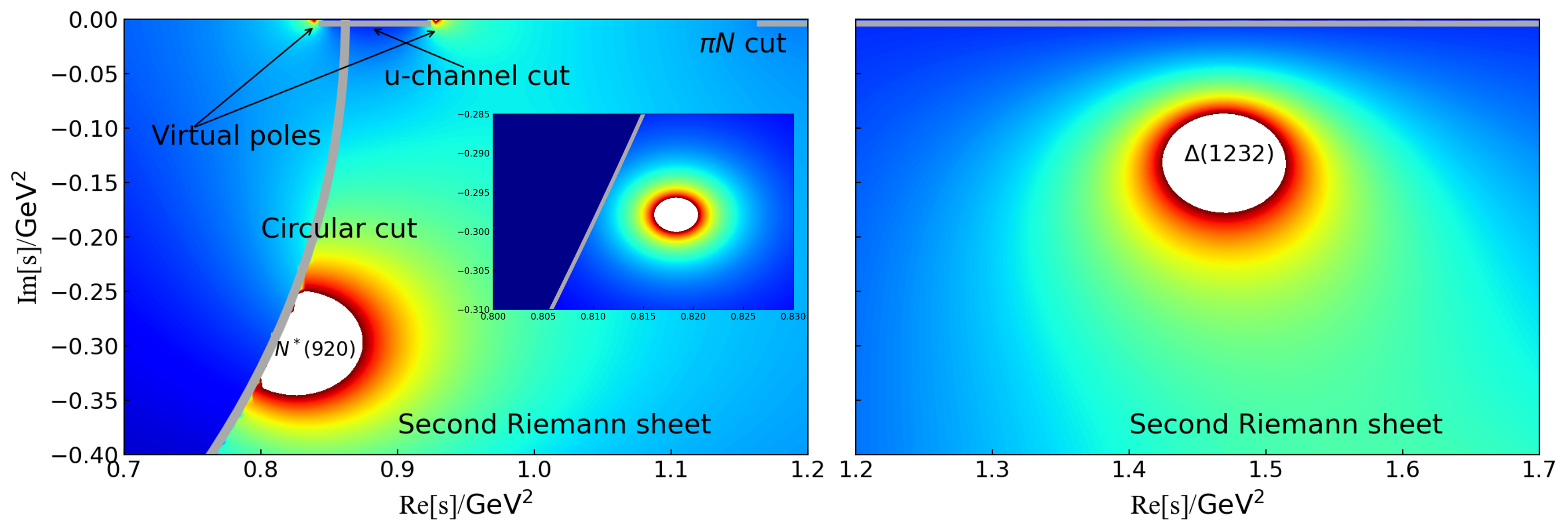}
    \caption{Left:$|S^{1/2~\mathrm{II}}_{0+}|$ on the $s$ plane, where $N^*(920)$ pole is clearly visible;
    Right: $|S^{3/2~\mathrm{II}}_{1+}|$ including $\Delta(1232)$ pole.}\label{fig:pole_920_1232}
\end{figure}

In order to determine the existence of the subthreshold pole, one must recast the S matrix as,
\begin{align}
    S^{1/2}_{0+}(s)=1-\frac{\sqrt{(s_{+}-s)(s-s_{-})}}{\sqrt{s}} f^{1/2}_{0+}(\sqrt{s})\ .
\end{align}
Using unitarity relation, $S^{1/2~\mathrm{II}}_{0+}=1/S^{1/2}_{0+}$, a pole on the second sheet of the S matrix, $S^{1/2~\mathrm{II}}_{0+}$, corresponds to a zero on the physical sheet. 
So all we need to do is to evaluate Eq.~(\ref{eq:f}) for complex values of $s$ in the validity domain and find out weather or not $S^{1/2}_{0+}$ has zeros there.  
Calculating $S^{1/2}_{0+}$ from the RS equation described above for the experimental inputs, we find that it does contain a zero, $S^{1/2}_{0+}(s_{N^*})=0$, with $\sqrt{s_{N^*}}=(0.918-i0.163)~\mathrm{GeV}$ (Herewith we denote the pole as $N^*(920)$).
In fact, we also find a zero in $S^{3/2}_{1+}$ (i.e. $P_{33}$ wave), $\sqrt{s_\Delta}=(1.210-i 0.047)~\mathrm{GeV}$~\footnote{As said before, at $\sqrt{s}=W_{\text{m}}$ the input phase shifts are taken from solutions of Ref.~\cite{Hoferichter:2015hva}. If we use instead solutions from GWU/SAID, we get $\sqrt{s_{N^*}}=(0.919-i0.162)~\mathrm{GeV}, \sqrt{s_\Delta}=(1.213-i 0.050)~\mathrm{GeV}$. The difference mainly comes from the fact that in $P_{33}$ channel, the GWU/SAID solution provides a phase shift about two degrees smaller than that of \cite{Hoferichter:2015hva}.}.
The second pole represents the well-known $\Delta(1232)$~\cite{Workman:2022ynf}.
According to our results, the existence of the wanted $S_{11}$ pole is established on the same footing as that of $\Delta(1232)$.
These singularity structures from the RS equations are depicted in Fig.~\ref{fig:pole_920_1232}.

As shown in Fig.~\ref{fig:pole_920_1232}, it is interesting that there are two virtual poles located at the neighborhood of short $u$-channel (nucleon) cut.
This general phenomenon was firstly discussed in \cite{PhysRev.123.692} (rediscovered in $\pi\pi$ scatterings~\cite{Zhou:2004ms}, in $\pi N$ scatterings~\cite{Li:2021oou}).
Taking for example $S^{1/2}_{0+}$ for discussion, it is real in the gap between the threshold $s_+$, and short $u$-channel cut. 
Since there are no bound states, $S^{1/2}_{0+}$ is bounded in the gap. Further, if there is no anomalous threshold, $S^{1/2}_{0+}$ is unity at $s_+$.
In \cite{Li:2021oou}, it has been proved that $S^{1/2}_{0+}$ approaches negative infinity when $s$ gets close to the short $u$-channel cut.
It is obvious in any case that the S matrix must have at least one zero on the first sheet, and therefore a pole on the second one. However, these two virtual poles hardly affect physical observables such as phase shifts~\cite{Li:2021oou}, in the scheme of the PKU decomposition of phase shifts~\cite{Zheng:2003rw, Zhou:2006wm, Yao:2020bxx}.
The $P_{11}$ channel is somewhat peculiar, due to the existence of nucleon bound state pole, it is impossible to obtain a correct phase shift without such a virtual pole~\cite{Wang:2017agd}.

Furthermore, the error bars of the pole positions can be estimated.
Unfortunately GWU/SAID data do not contain available error estimation, and the error bars from ref.~\cite{Hoferichter:2015hva} are also very small except that the two solutions differ in $P_{33}$ channel by about two degrees. 
In order to estimate it, we try to change GWU/SAID data to old KH80 results~\cite{Hohler:1983} to estimate the errors from matching phase $\delta^{1/2}_{0+}(1.36~\mathrm{GeV})$ and driving terms.
Because the $S$ wave scattering lengths are precisely determined irrespective of the uncertainty from driving terms, the noise in the scattering lengths is negligible for the pole position.
The location of the pole, $\sqrt{s_{N^*}}$, is therefore estimated to be
\begin{align}
    \sqrt{s_{N^*}} = (918 \pm 3)-i(163 \pm 9) \mathrm{MeV}\ .
\end{align}
As stated in Fig.~\ref{fig:fixb_PiN}, the point, $s_{N^*}$ within errors, is located inside the validity domain but outside the circular cut.
The simple estimation of errors leads to a rather small value.
For completeness, we have also given the estimation on $\Delta(1232)$ pole, $\sqrt{s_{\Delta}}=(1210\pm 4)-i(47 \pm 5) ~\mathrm{MeV}$.
The error bars for $\Delta(1232)$ location is  comparable to those found by \emph{Review of Particle Physics}~\cite{Workman:2022ynf}, hence it implies the estimation on the error bars of the $N^*(920)$ pole location is reasonable.

Our results are obtained based on PW analyses, with poles appearing in corresponding PW$s$. 
Such analyses only for $S$ and $P$ waves miss global constraints imposed by Regge theory that connects PW$s$ through analyticity in the angular momentum plane~\cite{Martin:1970hmp, Collins:1977jy, Donnachie:2002en, Gribov:2003nw}.
The most important feature of hadron spectrum is that its Regge trajectories characterized by parity and signature of a family are approximately linear.  
This was shown by Chew-Frautschi plot~\cite{Chew:1962eu}.
Consider the parity partner of $N_\alpha$ trajectory~\footnote{More detailed descriptions and properties for Regge trajectories can be found in Refs.~\cite{Collins:1977jy, Huang:2010aro, JPAC:2018zjz}.} (parity $\mathcal{P}=+1$ and signature $\mathcal{S}=(-1)^{J-1 / 2}=+1$), $N_\beta$. 
It is noticed that $N^*(920)$ pole could be added in $N_\alpha$ trajectory, i. e., the family with $N^*(1675)$ and $N^*(2250)$~\cite{Workman:2022ynf}.
It is reminded, however the $N^*(920)$ pole should not be considered as a well-established resonance yet.

\section{Summary}

It is clearly 
demonstrated  the existence
of a subthrshold pole in $S_{11}$ channel of $\pi N$ scattering amplitude. Since the pole is rather far away from physical region, only the powerful tool of analyticity can fulfil the task on its determination. 
The next question naturally arise is why it is physically relevant meanwhile it is far from the physical region.
We simply point out that it provides a large positive phase shift, without it one can not reproduce the experimental phase shift~\cite{Wang:2017agd}, in the scheme of PKU decomposition of phase shifts.
Furthermore, the appearance of the broad $S_{11}$ pole could be related to the broad $\sigma/f_0(500)$ and $\kappa/K^*_0 (700)$, since, in the same way as for the
meson resonances, the $S_{11}$ chiral amplitude involves a zero on the first sheet.
But of course, there are still lots of work to be done~\cite{Pelaez:2015qba, Pelaez:2020gnd} to further understand the property of such a pole.
Finally, our result also suggests a possible universal phenomenon for the appearance of such a broad structure~\footnote{Similar suggestion was also made in Ref.~\cite{Zhou:2020moj} from a different point of view.}.

\begin{acknowledgments}
X.-H. Cao and Q.-Z. Li would like to thank De-Liang Yao for helpful discussions on solving the MO problem.
We would like to thank Yu-Fei Wang and Zhi-Guang Xiao for valuable discussions.
We also thank Ulf-G. Mei{\ss}ner for a careful reading of the
manuscript and critical remarks.
Moreover, we wish to thank Frank Steiner for communicating many interesting information on the history of the development of Steiner equation.
This work is supported in part by National Nature Science Foundations of China (NSFC) under contract No. 11975028 and No. 10925522.
X.-H. Cao and Q.-Z. Li contributed equally to this work.
\end{acknowledgments}

\end{document}